\documentstyle[12pt]{article}
\def\gev{{\rm \, Ge\kern-0.125em V}}
\begin{document}
\begin{titlepage}
\pagestyle{empty}
\baselineskip=21pt
\rightline{McGill 95--17}
\rightline{UMN-TH-1333/95}
\rightline{hep-ph/yymmddd}
\rightline{June 1995}
\vskip .2in
\begin{center}
{\large{\bf Towards a Singularity-Free Inflationary Universe?}}
\end{center}
\vskip .1in
\begin{center}
Nemanja Kaloper

{\it Department of Physics, McGill University}

{\it Montr\'eal, Qu\'ebec, Canada H3A 2T8}

and

Richard Madden and Keith A. Olive

{\it School of Physics and Astronomy, University of Minnesota}

{\it Minneapolis, MN 55455, USA}

\vskip .1in

\end{center}
\vskip .5in
\centerline{ {\bf Abstract} }
\baselineskip=18pt
We consider the problem of constructing a non-singular inflationary universe
in stringy gravity via branch changing, from a previously superexponentially
expanding phase to an FRW-like phase. Our approach is based on the phase
space analysis of the dynamics, and we obtain a no-go theorem which
rules out the efficient scenario of branch changing catalyzed by dilaton
potential and stringy fluid sources. We furthermore consider the
effects of string-loop corrections to the gravitational action
in the form recently suggested by Damour and Polyakov.
These corrections also fail to produce the desired branch change.
However, focusing on the possibility that these corrections
may decouple the dilaton, we deduce that they may lead to an inflationary
expansion in the presence of a cosmological constant, which asymptotically
approaches Einstein-deSitter solution.

\vskip.5in
\centerline{\it Submitted to Nuclear Physics {\bf B}}

\end{titlepage}
\baselineskip=18pt
{\newcommand{\la}{\mbox{\raisebox{-.6ex}{~$\stackrel{<}{\sim}$~}}}}
{\newcommand{\ga}{\mbox{\raisebox{-.6ex}{~$\stackrel{>}{\sim}$~}}}}
\def\beq{\begin{equation}}
\def\eeq{\end{equation}}
\section{Introduction}

The standard cosmological model, based on Einstein's General Relativity
and the Cosmological Principle, represents a coherent, elegant and
consistent picture of much of our current understanding and observations
of the Universe. In fact, it is in excellent agreement with the
data going as far back as the era of primordial nucleosynthesis.
However, there remain well known gaps in this account, believed to originate
mainly from the times prior to nucleosynthesis, and manifest in such
problems as the size of the Universe, its smoothness
and flatness on very large scales and the lack of it at smaller scales,
the absence of topological
relics, such as domain walls, cosmic strings and magnetic monopoles,
the whereabouts of the missing matter, the absence (or near absence in Planck
units) of a cosmological constant, and perhaps the greatest mystery of all,
the initial singularity itself. The majority of the former problems can be
elegantly dealt with, at least in principle, by the inflationary paradigm
\cite{infl},
postulating an era of accelerated expansion before the era of
nucleosynthesis. Nevertheless, there yet remains the task to construct
a concrete, plausible dynamical scenario which predicts such evolution,
in agreement with observations and free of internal inconsistencies.
To render the matters worse, even if
a viable scenario is eventually brought to the light
of the day, it will still fail to address the problem of singularity
without resorting to major modifications of the theory of gravity, as
demonstrated by the Hawking-Penrose singularity theorems \cite{sing}.

As a consequence, there has been much labor in recent years in the attempt to
find alternative theories of gravity, reflecting the silent consensus
that certain alterations of Einstein's theory are imminent. In most cases,
the route taken was minimalistic in philosophy, consisting of inventing
and incorporating changes to account for some of the phenomenologically
dictated mechanisms. This approach has been only partially satisfactory,
failing to produce a model capable of dealing comprehensively with the
shortcomings of Einstein's theory, which proved deserving of respect
even in its demise.

On the other hand, the ongoing quest for the unified theory of interactions
in Nature has finally produced a promising contender, which has withstood
theoretical scrutiny up to date. The advent of string theory \cite{string},
based on a fundamentally different assumption about the nature of matter,
has been supported mainly by the description of gravity in a manner equivalent
to other forces, and perhaps even more importantly, by the absence of some of
the obstacles encountered in the failed attempts to quantize General
Relativity. Thrusting from the initial success, many studies have sprung up
investigating gravitational aspects of string theory, and in particular the
early universe cosmology \cite{dilinf}-\cite{bv}.
The justification for this interest can be naturally
found in the fact that string theory, while claiming to unify
gravity with the other forces of nature, must give us the means to investigate
the regions where the standard model has failed to give satisfactory
answers. The aforementioned cosmological problems, especially the problem
of the initial singularity, fall precisely in this category.
In addition, results of these investigations should provide us
with ways to test the compatibility of string theory with the Nature.

While the early investigations of string cosmology have indicated the presence
of some of the coveted mechanisms to tackle the encountered problems, they
have also burdened us with a host of new difficulties. In a typical
cosmological setting, most of the advantages and the difficulties can be
attributed to the presence of a new field, the scalar dilaton, which comes
in response to the requirement of conformal invariance of string world-sheet
theories. The dilaton has been recognized as a natural candidate for the
inflaton \cite{dilinf,bg}, a weakly coupled scalar which
is a necessary ingredient of many a
generic inflationary scenario. However, in addition to preordaining its
existence, conformal invariance also dictates the form of the dilaton's
couplings to gravity and other matter fields. Specifically, all particle
coupling constants and masses (believed to come about via some
symmetry-breaking mechanism) acquire dependence on the dilaton expectation
value. This yields to obstacles in implementing most of the conventional
inflationary scenarios in string theory \cite{bg},
compelling us to resort to
more obscure  and often contrived
arguments, thus diminishing the overall appeal of the theory.
These obstacles are best summarized by saying that the dilaton tends to roll
a bit too eagerly, influencing other fields in the model in such ways, for
example, as to preclude some well-known solutions such
as de-Sitter \cite{bd,sko}, affect
nucleosynthesis rates by inducing time-dependence of particle masses and
coupling constants \cite{dp,co},
and give rise to scalar components of gravity, similar
to the fifth force \cite{fifth}. It is therefore important
to see if there are ways to
keep the dilaton in check, and prevent it from meddling in the dynamics at
late times. Usually this is done by endowing the dilaton with a
large mass, coming from a dilaton self-interaction with a highly curved
potential well.

Another approach has been proposed recently
by Damour and Polyakov \cite{dp}.
Their model represents a further generalization
of the effective string gravity action, motivated mainly by the previously
described need to decouple the dilaton. The model is based on the observation
that scalar-tensor tensor theories with the scalar couplings to other
fields given by functions with a minimum at a non-zero value naturally evolve
towards Einstein's theory, and the inference that such couplings can arise
in string theory when the higher genus corrections are accounted for.
The resulting effect is that the dilaton can be stopped even in the absence
of conventional mass term-generating self-interactions. Obviously, the
model could become even more appealing if it contains additional features
compatible with our expectations from cosmology.

A very different approach towards string-driven inflation, dubbed
the pre-Big-Bang
inflation, has been suggested \cite{gv,g}. It strives to induce inflation
deriving from genuinely stringy mechanisms, relying on the dilaton,
scale factor duality symmetry of string theory (not present in any other
model of gravity), and somewhat less well understood possibility of
existence of string-induced topology-changing solutions \cite{ck}.
In its simplest form, this proposal rests on the extension of a simple
power-law expanding, scalar field-dominated universe, to negative times by
string scale factor duality and time inversion. The scenario is defined
in the string world-sheet frame, where the two branches are characterized by
superexponential inflation for $t<0$ and a milder, power-law expansion
for $t>0$. If the jump (branch change) at $t =0$ can be made
smoothly from the superinflationary phase
to the power-law one, the dreaded singularity may be avoided
\cite{bv}.

In this paper we will examine the
possibilities of avoiding the cosmological singularity \`a la Brustein and
Veneziano \cite{bv}. We will first revisit the case of dilaton self-interaction
and stringy fluid sources, originally investigated
by these authors, and improve their conjecture that these terms cannot
provide for a successful graceful exit, by promoting it
to an exact no-go theorem. We will then address the possibility for branch
changing using the Damour-Polyakov model, as well as
investigate further inflationary
capabilities of this model.
We shall, however, generalize the action given by Damour and
Polyakov one step further, to allow
for a non-trivial dilaton potential associated either with supersymmetry
breaking or a stringy cosmological term coming from the target-space
central charge deficit. Such extensions have also been considered recently in
\cite{dv}. We will again arrive at the concrete proof of
the no-go theorem for a branch change induced by the higher genus
terms in the gravitational action.
We will also note that our extension of the
Damour-Polyakov model brings de-Sitter-like solutions back into the game,
as asymptotic attractors reached when the dilaton decouples, as was also
discussed in \cite{dv}.

\section{The Gravitational Action and The Higher-Genus Contributions}

There is a considerable amount of literature concerning the tree level
gravitational action in string theory and its expansion in the string tension
($\alpha'$) \cite{acts}.
To $O(\alpha')$ the
effective Lagrangian describing the dynamics of the modes of interest to us may
be written as
\begin{equation}
S = \int d^4x \sqrt{g} \Bigl\{\frac{R}{2}
 - \partial_{\mu} \phi
\partial^{\mu} \phi + \frac{\alpha'}{16} e^{-2 \phi}(\hat
R^2-F_{\mu\nu}F^{\mu\nu}) - e^{2 \phi} \Lambda(\phi)\Bigr\}
\label{1}
\end{equation}
in the Einstein conformal frame. In (\ref{1}),
$\phi$ is the dilaton and $\Lambda(\phi)$ is the dilaton
potential (we shall be most interested in the case where
$\Lambda(\phi)$ is a constant, but we retain the $\phi$ dependence
to discuss the results of \cite{bv}).
We are using units such that $8 \pi G_N = 1$.
We have here ignored the contribution of the
axion since it enters
the equations of motion in the form $\propto c/a(t)^6$ (where $a(t)$ is
the FRW scale factor and $c$ a constant of integration),
so during expansion an axion dominated universe can be expected to quickly
evolve to one dominated by the dilaton \cite{lindil2,ko,gp}. Indeed, exact
results \cite{clw} show that the axion is significant only
in the past of any of the cosmological branches, and that it can neither
facilitate inflation nor provide for the solution of the graceful exit
problem.
Moreover, in what follows we will ignore the corrections of the derivative
expansion of order $O(\alpha')$ and higher. Though these terms would be
expected to be important near the singularity, we neglect them now solely for
reasons of simplicity.

We will, however, be interested in higher genus corrections. For this reason,
we reexpress our action, truncated to contain only the metric and
dilaton-dependent terms up to two derivatives, in the string world-sheet frame,
using the conformal transformation
$g_{\mu\nu} \rightarrow \exp(-2\phi) g_{\mu\nu}$. The action then becomes
\beq
S = \int d^4x \sqrt{g} e^{-2 \phi} \Bigl\{\frac{R}{2}
 + 2 \partial_{\mu} \phi
\partial^{\mu} \phi - \Lambda(\phi)\Bigr\}
\label{sact}
\eeq
Then, to parametrize the contributions from the higher genus terms to the
action, we make the Damour-Polyakov ans\"atz \cite{dp}
\beq
S = \int d^4x\,{\sqrt g}\,\left( B_g(\phi) R/2 +
2\,B_{\phi}(\phi)\left( \Box \phi - {{(\nabla \phi)}^2} \right)
- B_\Lambda(\phi) \Lambda(\phi) \right)
\label{act}
\eeq
The couplings $B_i(\phi)$ are
the functions accounting for the loop corrections and should
admit a weak coupling expansion of the form:
\beq
B_i(\phi)=e^{-2 \phi}+c_0^{(i)}+c_1^{(i)} e^{2 \phi}+c_2^{(i)} e^{4 \phi}
+\ldots
\label{bform}
\eeq
realized in the limit where $\phi \rightarrow -\infty$.

These corrections were examined
in \cite{dp} where the authors point out that at the tree level
the dilaton couples universally to all terms in the action
($c_j^{(i)}=0$ for all $i$ and $j$). Extending this presumption
of universality to the loop expansion (taking $B_i(\phi)=B(\phi)$
for all $i$) and conformally rescaling to
the Einstein frame results in mass dependencies of the fields
of the form $m_A(\phi)=m_A(B(\phi))$. Since the equations of
motion of the dilaton allow it to be attracted to extrema of
the mass functions as the universe passes through mass thresholds,
and all mass functions will generically have extrema coincident with those
of $B(\phi)$, the authors conclude that the dilaton
can be naturally decoupled in one such extremum of $B(\phi)$.
In light of this, we adopt the following action
\beq
S = \int d^4x\,{\sqrt g}\,B(\phi)\,\left( R/2 +
2\,\left( \Box \phi - {{(\nabla \phi)}^2} \right)
- \Lambda(\phi) \right)
\label{hgcc}
\eeq

The decoupling of the dilaton is necessary for a realistic
cosmology. A rolling dilaton will result in the time variation
in particle masses and gauge coupling constants. Their variations
can be sharply constrained by the requirement that they must
not disturb the delicate agreement between observed light element
abundances and their calculated values from the era of primordial
nucleosynthesis \cite{co}. It is known that the dilaton
does in fact decouple in a radiation dominated universe \cite{ko,tsey}.
On the other hand, in a matter dominated universe, at the tree level in the
gravitational action without a dilaton potential, the dilaton rolls \cite{ko}.
As mentioned above, in \cite{dp}, it was shown that higher genus corrections
of the form given in eq. (\ref{act}) can decouple the dilaton
without a dilaton self-interaction potential.
It is also known that the dilaton must decouple in order for the universe
to achieve a deSitter like expansion \cite{bd,sko}, ie. simply the presence
of a cosmological constant would not bring about exponential expansion
as (at the tree level) in the Einstein frame the cosmological term would
carry a factor of $e^{2\phi}$.  While a potential for the dilaton which
is expected to be generated by supersymmetry breaking can trap the
dilaton leading to exponential expansion \cite{sko,ko}, below we will
consider the possibility that the higher genus terms in the action play a
similar role, thus allowing for inflation without a dilaton potential.

We choose to extract the equations of motion directly in
the string frame, since here we will be able to express
them in terms of a first-order system of non-linear
differential equations and apply dynamical methods
to analyze the evolution.
We also specialize to the case of a Friedmann-Robertson-Walker
cosmology and take a metric of the form:
\beq
ds^2=-n(t)^2 dt^2+a(t)^2 d{\vec x}^2
\eeq
where $d{\vec x}^2$ is the three dimensional volume element for a space of
constant curvature $k$. We will concentrate on the case
of $k=0$. Expressing the contents of the action in terms of
the functions $a(t)$ and $n(t)$ we can factor out the
spatial integration and perform the variation with respect
to the functions $a(t)$, $n(t)$ and $\phi(t)$, and finally
set $n(t)=1$. The equations
will contain $B'(\phi)$ and $B''(\phi)$. In view of the
expansion (\ref{bform}) it will be convenient to define
$\beta(\phi)=B'(\phi)/B(\phi)+2$ and therefore
$B''(\phi)/B(\phi)=\beta'(\phi)+(\beta(\phi)-2)^2$ so
that the weakly coupled regime $\phi \rightarrow -\infty$
corresponds to $\beta \rightarrow 0$. The
equations of motion for the tree level in the string loop
expansion ($c_0,c_1,c_2 \ldots = 0$) thus correspond to $\beta=0$.

The resulting equations of motion are:
\begin{eqnarray}
0&=&3\,{h^2} - \Lambda(\phi) + 2\,{\dot h} +
2\,(\beta(\phi)-2)\,h\,{\dot \phi} + \nonumber \\
&{}& (2-2\,\beta(\phi) +
{{\beta(\phi)}^2}+\beta'(\phi))\,{{{\dot \phi}}^2} +
(\beta(\phi)-2)\,{\ddot \phi} \label{e1}
\\
\noalign{\medskip}
0 &=& 6\,(\beta(\phi)-2)\,{h^2} +
(2-\beta(\phi))\,\Lambda(\phi) +
3\,(\beta(\phi)-2)\,{\dot h} - \Lambda'(\phi) +
\nonumber \\
 & &12\,(1-\beta(\phi))\,h\,{\dot \phi} +
2\,(-2+3\,\beta(\phi)-\,{{\beta(\phi)}^2}-
\,\beta'(\phi))\,{{{\dot \phi}}^2} + \nonumber \\
 & &4\,(1-\beta(\phi))\,{\ddot \phi} \label{e2}
\\
\noalign{\medskip}
0 &=& 3\,{h^2} - \Lambda(\phi) + 3\,(\beta(\phi) - 2)\,h\,{\dot \phi}
+ 2\,(1-\beta(\phi))\,{{{\dot \phi}}^2} \label{e3}
\end{eqnarray}
(\ref{e1}) results from the variation of the scale
factor $a(t)$, (\ref{e2}) from the variation of dilaton $\phi(t)$ and
(\ref{e3}) from the variation of the lapse factor $n(t)$.

\section{Branch Changing and the Graceful Exit}

Before we begin our analysis, we should
review the main ingredients of solution to the graceful exit
problem proposed by Brustein and Veneziano \cite{bv}. This possibility
was originally described in the
genus-zero $O(\alpha'^0)$ approximation,
with some dilaton potential $\Lambda(\phi)$, where the dynamics is defined
by the action (\ref{sact}).
Generalization
to the case when stringy fluid is present is straightforward, and we will
reflect on it later.
To recover the corresponding equations of motion
we set $\beta(\phi)=0$ in (\ref{e1}-\ref{e3}).
It was noted  \cite{bv} that the resulting equations of motion can be
solved explicitly for $\dot \phi$ and  $\dot h$. The canonical first order
system takes the relatively simple form:
\begin{eqnarray}
\dot \phi &=& (3 h \pm \sqrt{3 h^2+2 \Lambda(\phi)})/2 \label{bv1} \\
\noalign{\medskip}
\dot h   &=& \pm h \sqrt{3 h^2+2 \Lambda(\phi)}-\Lambda'(\phi)/2
\label{bv2}
\end{eqnarray}
with the $\pm$ sign chosen for both equations simultaneously.
These equations are easily solved in the case $\Lambda=0$
resulting in four different solutions, two for each of the
two branches corresponding to the choice of
(+) and ($-$) sign. The (+) branch is defined in the
domain $t<0$ and the ($-$) branch in $t>0$.
The solutions are \cite{bv}
\begin{eqnarray}
&&a=a_0 |t|^{\mp \frac{1}{\sqrt{3}}}, ~~~ h = \mp \frac{1}{\sqrt{3} t},
 ~~~ \phi = \phi_0 + \frac{\mp \sqrt{3} - 1}{2} \ln |t|
{}~~~~ {\rm for} ~t<0 ~((+) ~{\rm branch}) \nonumber \\
&&a=a_0 t^{\pm \frac{1}{\sqrt{3}}}, ~~~~~ h = \pm \frac{1}{\sqrt{3} t},
 ~~~ \phi = \phi_0 + \frac{\pm \sqrt{3} - 1}{2} \ln t
{}~~~~~ {\rm for} ~t>0 ~((-) ~{\rm branch}) ~~~~~~
\label{bvsols}
\end{eqnarray}

Note that the expanding solution for $t<0$
begins in the weak coupling
regime ($\phi$ large and negative) and evolves
toward the strong coupling region
($\phi$ positive), compatible with the form of the action which
represents a weak-coupling truncation of the full effective
action of string theory. In the following, we will
want to choose the expanding solution for $t>0$
as we are motivated by a desire to match this solution
with a Robertson-Walker
cosmology having decelerated expansion.
In contrast, the contracting
solution for $t<0$, begins in the strong coupling region and evolves towards
the weakly coupled region of field space. In some sense, unless string loops
are taken into account, this truncation is somewhat ad hoc for the
contracting solutions.
Now,
if we look at the time evolution of the scale factor,
we can have either expansion or contraction at $t<0$, yet
we are interested in expansion only at $t>0$.
For the
expanding/expanding (contracting/expanding)
solutions we note that for $t<0$ we have a pole-driven
expansion (contraction), reaching the singularity at $t=0$,
and for $t>0$ we get a power-law
expanding universe emerging from the singularity.

If these two temporal branches
could be viewed as a single solution, ignoring the presence of the curvature
singularity at $t=0$ for the moment, the compound
configuration could perhaps possess quite
remarkable properties, carrying out most of the commandments of the
inflationary doctrine. Namely, prior to the instant $t=0$ we'd see a
superexponential, pole-driven inflation.
The inflation would be driven solely by the dilaton
kinetic energy, thus doing away with the need for more complicated sources.
Moreover, if the singularity at the pole $t=0$ can
be surmounted and the two temporal branches joined smoothly,
the resulting solution would represent a completely nonsingular cosmology.
As the curvature approaches the
Planck scale, from a prior expanding or contracting solution,
to eventually exit this region and metamorphose to a cooling,
expanding universe which can be joined onto our own
as $t \rightarrow \infty$ \cite{bv}. The bump around $t=0$ would then
resemble the Big Bang and therefore, in addition to possibly solving the
problems usually assigned to inflation, it would also give an elegant
resolution to the question of initial singularity.

In general, we will see that the asymptotic properties of the
(+) and ($-$) branches will require that we switch from one
to the other to stay within the limits of our theory and
to get a desirable late time cosmology.
These properties can be summarized in the observation that
(+) branch solutions evolve towards singularities in
their future while ($-$) branch evolve away from singularities
in their past. It is obvious from the equations of motion that the two branches
can never connect smoothly in the regions where the
potential is positive (cf. eqs. (\ref{bv1})
and (\ref{bv2})). Namely, if the two
branches are to be continuously attached to each other, at the location of
contact the values of derivatives must be the same. This requires that
the potential be negative in a certain region.
If we represent the dynamics by the phase space portrait in the phase plane
$(\phi, h)$, the regions where branch changes can occur are closed
curves symmetric around the $\phi$-axis, given by $3h^2 +2\Lambda = 0$.
They were conveniently named the ``eggs" because of their
concave shape in the regions containing a single negative minimum of
$\Lambda$.

Before considering whether such successful branch changes can be catalyzed
by eggs in general, we will present here the
special case when
$\Lambda(\phi)=const$. We look at this case for two reasons.
The case $\Lambda<0$
gives us the simplest example of a potential with an egg - in fact, with
nothing else but the egg, because the potential is negative everywhere, and
thus the egg is just two lines parallel with the $\phi$-axis.
The other case, $\Lambda>0$, has no eggs, but it gives us a
clear description of the generic properties of solutions in the
regions of fairly flat potentials, and allows us to identify
the associated attractors and repellers as the linear dilaton
vacua, which is a well understood conformal field theory construction.
This shows that we can think of the linear dilaton vacua as seeds for (+)
branch superexponential inflation.
Furthermore, in these two cases the equations of motion
(\ref{bv1}) and (\ref{bv2})
can be integrated exactly, and we can use the solutions to develop our grasp
of the qualitative properties of solutions in more general cases.

We present only the solutions where
$h>0$ since the $h<0$ solutions may be obtained from these by time-reversal
$t \rightarrow -t$ (changing $h \rightarrow -h$).
In the case $\Lambda < 0$ we obtain:
\begin{eqnarray}
h &=& \sqrt{2|\Lambda|} / (\sqrt{3} \sin ({\sqrt{2|\Lambda|} t})) \\
\noalign{\medskip}
\phi &=& (1/2) (\sqrt{3} \ln (\tan({\sqrt{2|\Lambda|} t}/2))
- \ln (\sin({\sqrt{2|\Lambda|} t}))) + \phi_0
\end{eqnarray}
where $0<t<\pi / {\sqrt{2|\Lambda|}}$.
These solutions feature
a branch change (from ($-$) to (+) at $t=\pi /(2 \sqrt{2|\Lambda|})$
in analytic form.
To see this,
notice the sign change in $\dot h$ and recall the presence of the upper
boundary of the egg (the upper of the lines
$\left| h \right|=\sqrt{2|\Lambda|}/\sqrt{3}$).
Notice that the single trajectory is this case is singular at both endpoints.
(The simple identification of the (+) branch being associated with negative
times naturally followed by a ($-$) branch at positive times is no longer
convenient when a potential is present. As always, the (+) branch evolves
towards a singularity and the ($-$) branch away from one.)
Qualitatively, we see that the picture is as follows: the universe
begins on a ($-$)
branch near the Big-Bang singularity, which in the phase space approach
corresponds to the limit $\phi \rightarrow -\infty$, $h \rightarrow \infty$.
The universe then evolves along the ($-$) branch towards the strong coupling
regime, with its expansion being decelerated. Eventually, the evolution brings
it down near the upper egg line, which it touches tangentially, and moves away
on the (+) branch. This branch goes steadily upwards, never returning to the
vicinity of the egg line, and thus we end up with a branch change in the
direction reverse to the one we are looking for. An example is shown in
Fig. (1).

In the case $\Lambda>0$ there is no egg,  trajectories from different
branches never link up. One can see that now there exist two classes of
solutions. The phase trajectories with variable Hubble parameter are
directly analogous to the two $\Lambda=0$ branches.
The ($-$) branch is given by:
\begin{eqnarray}
h &=& \sqrt{2\Lambda} / (\sqrt{3} \sinh ({\sqrt{2\Lambda} t})) \\
\noalign{\medskip}
\phi &=& (1/2) (\sqrt{3} \ln (\tanh({\sqrt{2\Lambda} t}/2))
- \ln (\sinh({\sqrt{2\Lambda} t}))) +
\phi_0
\end{eqnarray}
where $t>0$.
Similarly, the (+) branch solution is given by:
\newpage
\begin{eqnarray}
h &=& \sqrt{2\Lambda} / (\sqrt{3} \sinh ({-\sqrt{2\Lambda} t})) \\
\noalign{\medskip}
\phi &=& (1/2) (-\sqrt{3} \ln (\tanh({-\sqrt{2\Lambda} t}/2))
- \ln (\sinh({-\sqrt{2\Lambda} t}))) +
\phi_0
\end{eqnarray}
where $t<0$.
In addition there are the linear dilaton vacua themselves,
given by \cite{lindil1,lindil2}
\begin{equation}
h = 0 ~~~~~~~~~~~~ \phi = \phi_0 \pm \sqrt{\Lambda/2} ~~t
\end{equation}
They generalize the trivial solutions $h=0, \phi={\rm const}$ present in the
$\Lambda=0$ case.
The linear dilaton vacuum solutions do not appear when $\Lambda <0$, as
$\dot \phi$ would become imaginary.
{}From comparing these solutions we see that the ($-$) branch
solutions emerge from the singularity at $t=0^{+}$
and approach the linear dilaton vacuum with the $-$ sign in the above
equation as $t \rightarrow \infty$.
For the (+) branch case, the solutions begin at the linear dilaton vacuum
with the positive sign chosen in the equation above
as $t \rightarrow -\infty$, and evolve towards the singularity as
$t \rightarrow 0^{-}$.
This identifies for us the asymptotic conditions
for the solutions with negligible
potential gradients. Generically, the (+) branch solutions evolve away from
linear dilaton vacua (even in the case without the cosmological constant,
which we can think of as $\Lambda = \epsilon^+ \rightarrow 0$), and ($-$)
evolve towards the linear dilaton vacua. In particular, this singles out a
specific initial condition for the universe which starts on a (+) branch,
and thus does away with the initial condition problem, as we have indicated
above. These solutions are graphically represented in Figs. (2) and (3).

The family of solutions for constant $\Lambda$ (the non-zero central charge
deficit)  may
also be found implicitly in \cite{gp}. There the effects of the axion term
were included too. The above solutions with $\Lambda = const$ may be
seen at the boundary of their
figures in the $(\dot \phi,h)$ plane where the axion goes to zero.

Armed with these examples and intuitive arguments, we can delineate the
properties of a non-singular cosmology and simultaneously with it the
properties of a potential that would guide its evolution.
To avoid all singularities, we must have a branch change,
so we require a potential to become negative,
producing one or more eggs.  Given
such a favorable potential, in investigating the possibility of a branch
change from (+) to ($-$), Brustein and Veneziano arrived at
the ``graceful exit'' problem. Based on numerical integration
of the equations of motion augmented
with some qualitative arguments,
they concluded that while in all the
cases they have analyzed it was possible to
induce a change (+) $\rightarrow (-)$, it was always followed by
another change $(-) \rightarrow (+)$, and the problem persisted.
Due to the approximate nature of these arguments, they
referred to their results as a ``very-hard-to-go-theorem".
In fact, an exact result can be obtained. We will present
this no-go theorem in the next section.

With the failure of the dilaton potential alone to produce the
required branch changing, the authors of \cite{bv} attempted improve
the situation by including
stringy fluid sources, higher dimensional embeddings, and combinations
of all of them without success. After presenting the proof of the exact
no-go theorem for dilaton potential, we will generalize it by outlining
the proof when the fluid sources are present.

It has been advocated that the singular behavior
of the cosmological solutions may be
resolved with the help of higher derivative terms, important in the regions
of large curvature, which have been shown to lead to interchanges of
duality-related branches in asymptotically weakly coupled, flat regions
\cite{ck}. This is further supported by the existence of completely
non-singular, non-perturbative cosmologies in string-like models with the
dilaton coupled to the Gauss-Bonnet higher derivative curvature combination
\cite{art}. We underline here that all the nonsingular solutions presented in
\cite{art} are nonperturbative in the strength of the Gauss-Bonnet coupling,
which one should expect on the grounds that there are no nonsingular FRW
cosmologies in the absence of this term. This points to the
fact that these solutions cannot be immediately regarded as string
cosmologies, because other higher order corrections may be important.
Concrete information is still lacking, however, due to the absence of a
general procedure to treat the higher order corrections to all orders
in a systematic way, distinguishing between the physically
relevant contributions and counterterms arising due to the redefinition
ambiguity. In the absence of this, we feel that it is of interest to look at
other options and attempt to clarify the essentials of this graceful exit
problem.

\section{The Proof of The No-Go Theorem}

In this section we will consider the problem of potential-catalyzed
branch changing and prove that it cannot occur in the scenario
envisioned in the previous section.
This therefore rules out potentials as possible solutions of the
graceful exit problem in stringy cosmology. We will also show that it is
straightforward to generalize this result to the case when stringy fluids are
present.

The main thrust of this argument will concern the behavior of solutions
which bounce off the egg, which we will show can't lead to favorable
branch changes. But before we embark on this, we remark that
numerical experiments show that it
is extremely difficult to get a `good' (+) initial trajectory
(in the sense of arising from no past singularity)
to hit an egg at all for a simple looking potential.
The reason for this is the presence of
a saddle point in the flow ahead of the egg which divides
the `good' trajectories into two streams which flow around the
egg. In fact, we can show analytically that no such trajectories can hit
the egg generated by a positive curvature quadratic potential.
But since we can't rule out the possibility of a first touch
on the egg of a sufficiently bizarre shape, we must consider
the possibility of a bounce. In this case, we \underbar{stress} that
the proof applies to \underbar{any} (+) trajectory originating outside of
the egg region.

To begin the inquiry into the global properties of the
solutions for a general $\Lambda(\phi)$ we shall find the loci of points
where the direction of flow of trajectories in the phase plane changes.
These are given by the curves where $\dot h=0$ and $\dot \phi=0$.
The detailed properties of these curves are not important
to our conclusions, and we will state merely those which we
require as we quote the results.
If we solve (\ref{bv2}) with the condition
$\dot h=0$, we find:
\begin{eqnarray}
&&h^2 = (1/6) (-\Lambda(\phi)+
\sqrt{\Lambda(\phi)^2+(3/4)\Lambda'(\phi)^2}) \nonumber \\
&&h\Lambda' \ge 0 ~~ {\rm for~(+)~branch} ~~~~~~~
h\Lambda' \le 0 ~~ {\rm for~(-)~branch}
\label{hdot0}
\end{eqnarray}
By directly examining (\ref{bv2}) for large positive and
negative values of $h$, we notice that for the (+) branch,
the $h$-flow is away from the
curve ($\dot h>0$ above it and $\dot h<0$ below),
whereas for the ($-$) it is towards it ($\dot h<0$ above
the curve and $\dot h>0$
below it). This behavior justifies our characterization of (+)
trajectories as singular in the future (positive feedback)
and ($-$) branch as
singular in the past (negative feedback).

If we solve (\ref{bv1}) with $\dot \phi=0$, we obtain two
curves where the flow of $\phi$ changes sign:
\begin{eqnarray}
&&h^2 = \Lambda(\phi)/3 \nonumber \\
&&h \le 0 ~~ {\rm for~(+)~branch} ~~~~~~~
h \ge 0 ~~ {\rm for~(-)~branch}
\label{pdot0}
\end{eqnarray}
Notice that these curves touch the vertical strip containing the
egg only at the very ends (where $\Lambda(\phi)=0$) and extend
away from the egg region. Thus, the $\phi$ flow vertically
above the egg is from left to right
(from the weak coupling towards the strong coupling)
and reversed below them, for both branches. Putting these facts
together we see that trajectories tend to flow clockwise
around the egg.

At the intersection of these curves lie fixed points. The
conditions for these are most easily read off by setting
the dotted quantities to zero in the second order equations
\begin{eqnarray}
&&3h^2=\Lambda(\phi) = -\Lambda'(\phi)/2 \ge 0 \nonumber \\
&&h \le 0 ~~ {\rm for~(+)~branch} ~~~~~~~
h \ge 0 ~~ {\rm for~(-)~branch}
\label{hpdot0}
\end{eqnarray}
These fixed points come in pairs, above the $\phi$-axis for the ($-$) branch
and below it for (+), except when they coincide for the cases when $h=0$.
We can analyze the nature of these fixed points directly in the
string frame and determine that if the quantity $\Lambda'(\phi)+
\Lambda''(\phi)/2$ is negative, the points are hyperbolic
(saddle points) and where it is non-negative the ($-$) branch
$h \ge 0$ is an attractor and the (+) branch $h \le 0$ point is a
repeller. None of this should come as a surprise, since thinking
of these fixed points in terms of the potential in the Einstein
frame $e^{2 \phi} \Lambda(\phi)$, we see that the saddle points
correspond to positive maxima and the attractor/repeller pairs
to positive minima of this potential. This correspondence is
exact since near the fixed points $\phi$ is moving very slowly, and
the conformal rescaling between the frames is nearly
constant, so that the notion of the character of a fixed
point does not depend on the frame.

Following \cite{bv} we define the egg function:
\begin{eqnarray}
e=\sqrt{3h^2+2 \Lambda(\phi)}
\label{egg}
\end{eqnarray}
Then it is easy to show that
\begin{eqnarray}
\dot e=\pm (1/2)(6h^2+\Lambda'(\phi))=\pm (2 h \dot \phi-\dot h)
\label{dote}
\end{eqnarray}
The condition $\dot e =0$ defines curves separating regions
where the egg attracts or repels different branches and
extends the infinitesimal condition in \cite{bv} governing approach to
the egg. Of more interest than the curves themselves is the fact that
dividing both sides of (\ref{dote})
by $\dot \phi$ and integrating the result over $\phi$ along a trajectory,
we obtain:
\begin{eqnarray}
\pm(e(t_1)-e(t_0))+h(t_1)-h(t_0)=2 \int_{\phi(t_0)}^{\phi(t_1)} h d\phi
\label{diffint}
\end{eqnarray}
The integral here is to be understood as a line integral along the path
of the system between $\phi(t_0)$ and $\phi(t_1)$.
As we will now show, the sign of this integral
represents an exceptionally strong constraint on the behavior of trajectories,
and provides the needed tool to obtain the no-go theorem.

Let us now outline our proof. We should recall that under
a successful branch change we mean a trajectory which
enters the egg region as a (+), since they can have non-singular pasts,
and leaves on a ($-$) branch, since they have
have non-singular futures and indeed may be
captured by a ($-$) attractor if we wish to decouple the dilaton
(recall there are no (+) branch attractors).
We will show that such a successful branch change is impossible, and
will base the proof on three important details.
First, we reemphasize that \underbar{all} trajectories flow from left to
right in the region vertically above the egg and right to
left below. Thus any trajectory hitting the top of the
egg must come from the left and any trajectory hitting below
the egg must come from the right, if they are to have any extension
outside of the egg region. This, of course, allows that a trajectory can flow
around the egg without hitting it for ``half" a cycle, e.g. coming from the
right in the far past, flowing below the egg and reemerging above the
$\phi$ axis to the left of an egg, flowing towards it.
Second, we will prove that any ($-$) branch trajectory,
originating from anywhere on the upper side of the
egg cannot escape over the right end of the egg but must hit it again.
The third ingredient of our proof is a time-reversed corollary of the
second, that any (+) trajectory coming from the right and flowing below the
egg cannot hit the egg below, or on,
the $\phi$-axis. These latter two impossible trajectories are
illustrated in fig.~(4).

Combining these together we see that
any (+) branch entering an egg region from the left must go over the top
of the egg, possibly experiencing several branch changes,
and must exit the region of the egg to the right still being
on the (+) branch. Any (+) branch entering an egg region from the right
is prohibited from hitting below, and so it must remain (+) while flowing
under the egg. Thus any (+) trajectory entering the egg region
cannot leave on a ($-$) branch, and there
is no graceful exit. The egg can only convert
($-$) to (+). Clearly, multiple eggs cannot
change this conclusion.

Now it remains to establish the second and third of our claims.
First we show that a ($-$) branch bounce, originating from anywhere on
the upper side of the egg cannot escape over the right end
of the egg but must fall down on it again. To see this recall the integral
formula (\ref{diffint}) for a ($-$) trajectory:
\begin{eqnarray}
-e(t_1)+e(t_0)+h(t_1)-h(t_0)=2 \int_{\phi(t_0)}^{\phi(t_1)} h d\phi
\label{diffint-}
\end{eqnarray}
Let $t_0$ be the time of the origin of the ($-$) bounce, and
$t_1$ the later time when the trajectory leaves
the end of the egg.
Then, at the egg $e(t_0)=0$, and $h(t_0) \ge 0$.
At the end of the egg $h(t_1) \ge 0$ and $e(t_1) = \sqrt{3} h(t_1)$,
since the end of the egg is defined by the condition
$\Lambda(\phi(t_1))=0$.
Finally, in this region the flow is to the right and $h \ge 0$.
Therefore, the integral is equal to the area between
the segment and the $\phi$-axis and hence strictly positive
(we will remark below on the degenerate cases where this area
may be zero):
\begin{equation}
\int_{\phi(t_0)}^{\phi(t_1)} h d\phi = {\cal A} >0
\label{area1}
\end{equation}
Substituting these in (\ref{diffint-}) we arrive at the sought
contradiction:
\begin{equation}
0<2{\cal A}=-(\sqrt{3} -1) h(t_1) - h(t_0) \le 0
\label{contra1}
\end{equation}
Therefore, the ($-$) bounce emerging from the upper side of the egg must
terminate back on it, as we claimed.

For the third claim
consider a (+) branch entering the region below the egg and
passing the right end of the egg.
Recall the integral
formula (\ref{diffint}) for a ($+$) trajectory:
\begin{eqnarray}
e(t_1)-e(t_0)+h(t_1)-h(t_0)=2 \int_{\phi(t_0)}^{\phi(t_1)} h d\phi
\label{diffint+}
\end{eqnarray}
Let $t_0$ be the time of passing the end of the egg, defined
by the condition that $h(t_0) \le 0$ and $\Lambda(\phi(t_0))=0$,
so that $e(t_0)=\sqrt{3} | h(t_0) |$.
Let $t_1$ the later time when the trajectory hits below the
egg, so that $e(t_1)=0$ and $h(t_1) \le 0$.
In this region the flow is to the left and $h \le 0$
so the integral is again equal to the area between
the segment and the $\phi$-axis and strictly positive:
\begin{equation}
\int_{\phi(t_0)}^{\phi(t_1)} h d\phi = {\cal A} >0
\label{area2}
\end{equation}
Substituting these in (\ref{diffint+}) we arrive at the sought
contradiction:
\begin{equation}
0<2{\cal A}=-(\sqrt{3} -1) | h(t_0) | - | h(t_1) | \le 0
\label{contra2}
\end{equation}
This contradiction shows the (+) trajectory cannot hit below
the egg.

This concludes the main line of argument for the ``no-go'' theorem.
However, several exceptional cases remain which we will deal with
briefly. While these cases require an infinite fine
tuning of initial conditions, we will show they
can be dismissed. One may worry about the case of tangential hits on the
very ends of the egg where it meets the $\phi$ axis, or possibly
passes through ``pinches'' where the egg narrows to a single
point. If this happens at the left end of the egg, where
generically $\Lambda'(\phi)<0$, we can substitute $h=\Lambda(\phi)=0$
into the equations of motion (\ref{e1}), (\ref{e2}) and
(\ref{e3}) to conclude the $\dot \phi=0$ and $\dot h=\ddot \phi=
-\Lambda'(\phi)/2>0$. Hence this point on the curve is a
minimum of $\phi(t)$, and the trajectory is curving from the
region vertically below the egg (+) into the region
vertically above the egg ($-$), and we
can easily see this cannot lead to any exceptional behavior.
A hit on the right point is a change from ($-$) above to (+)
below and again does not lead to exceptional cases.

Hits at inflection points where $h=\Lambda(\phi)=\Lambda'(\phi)=0$
(similar to a ``pinch'', but occurring at the ends
of the egg, or even more generally corresponding to a region where
$\Lambda(\phi)=0$ for an interval on the $\phi$ axis)
may seem more troublesome, since
it will be difficult to extract information about the
past and future of these trajectories. But here we can
refer to a general property of the second-order
equations of motion. We note that they can be written
in the form of a normal system \cite{odes}, i.e. that
the second derivatives $\ddot \phi$ and $\ddot h$ can be
written as functions of the first derivatives and
values of $\phi$ and $h$. If we require that these functions are
Lipschitz in their arguments in a neighborhood
of the point of interest, we may conclude that
the trajectories are unique for given initial
conditions there. (The Lipschitz condition is
a weaker form of a bounded derivative condition).
Since this is a natural local condition for
$\Lambda(\phi)$ and $\Lambda'(\phi)$ we conclude that
no two trajectories of the same branch can
intersect. Now to return to the inflection points,
we notice that they are also well-behaved fixed points, since there
is a trivial solution ($\phi$ and $h$ constant)
sitting in them, and thus no other solution can
cross through them, but only approach them asymptotically.
Therefore the inflexions cannot be used for
branch changing, as no bounces can originate from them.

In addition, one might wonder whether solutions can
circle and change branches on the egg ad infinitum, leading
to a curious quasi-cyclic cosmology. Briefly, the answer
is no. The equation (\ref{diffint}) can be used to show
that each hit on the egg must be higher than the
previous, and in fact must be at least twice the area of the
egg higher for each rotation about it. This result
can be sharpened to show that no incoming ``good'' (+)
can circle the egg and rehit it on the top. A ``zero area''
egg is no solution either, since this is a (+) branch
repeller. To see this we note that the motion around such a
point is clockwise and the accumulation of area in
the integral formula will push (+) branch solutions
away.

Finally, simply for completeness, we note the presence of
fine tuned (+) branch solutions whose evolution
asymptotically slows down to a halt at a saddle point at
$h < 0$, approaching a contracting deSitter phase.
With the exception of these, and the constant $h$ solutions sitting
at fixed points to which these solution tend, our arguments show that
all other evolutions must begin or end in singularities, or both.

As we have mentioned before, this no-go theorem can be generalized
to the case when stringy fluid sources are present. We will now outline
the proof for this case. We extend the phase space of the model to three
dimensions, the third coordinate being the energy density of the fluid
$\rho$. The associated equations of motion are given by the following
generalization of (\ref{bv1}-\ref{bv2}) \cite{bv}:
\begin{eqnarray}
\dot \phi &=& (3 h \pm \sqrt{3 h^2+2 \Lambda(\phi)
+ \rho \exp{(2\phi)}})/2 \nonumber \\
\noalign{\medskip}
\dot h   &=& \pm h \sqrt{3 h^2+2 \Lambda(\phi) + \rho \exp{(2\phi)}}
-\Lambda'(\phi)/2 + \frac{\gamma}{2} \rho \exp{(2\phi)} \nonumber \\
\noalign{\medskip}
\dot \rho &=& - 3(1 + \gamma) h \rho
\label{source}
\end{eqnarray}
Here $\gamma = p/\rho$, is a constant representing the
fluid equation of state, for which we will only require
$\gamma > -1/3$. This includes a wide range of fluids, both stringy  with
$\gamma \in (-1/3, 1/3)$ \cite{bv}, and relativistic, corrected by the
dilaton coupling as discussed in \cite{ko,dp} and references therein.
We note that the physical restriction $\rho \ge 0$
is consistent with the equations of motion, as the $\rho$ flow terminates at
the $\rho=0$ plane, which is like a potential barrier. Moreover, we note
that the trajectories completely confined in this plane are governed by our
previous theorem, so there is no graceful exit for them. Now we look at
the fully three-dimensional trajectories. The egg function is given by
\begin{equation}
e=\sqrt{3 h^2+2 \Lambda(\phi) + \rho \exp{(2\phi)}}
\label{fluidegg}
\end{equation}
Taking a time derivative of (\ref{fluidegg}), dividing the resulting equation
by $\dot \phi$ and integrating over $\phi$ along a trajectory, we obtain
the modified integral formula, analogous to (\ref{diffint}) when sources
are present:
\begin{eqnarray}
\pm(e(t_1)-e(t_0))+h(t_1)-h(t_0)=2 \int_{\phi(t_0)}^{\phi(t_1)} h d\phi
+ \frac{1+\gamma}{2} \int_{t_0}^{t_1} \rho e^{2\phi} dt
\label{diffintfl}
\end{eqnarray}
This equation differs from (\ref{diffint}) only in the presence of the last
term, which is a \underbar{nonnegative} \underbar{quantity}
for all trajectories,
as the energy density is restricted by $\rho \ge 0$. Now, in
this case the other relevant characteristics of the phase space, given by
equations (\ref{hdot0}-\ref{hpdot0}) are easily generalized to three
dimensions. We will not present the details here. It is sufficient to see
that qualitatively the picture remains the same: the egg is now a
two-dimensional compact surface cut by the plane $\rho=0$, and the only
fixed points on it may again be ``pinches" or inflexions in the $\rho=0$
plane. The flow of trajectories around the egg is generically along helical
paths, which if projected onto the $\rho=0$ plane turn clockwise. We then
need to consider trajectories crossing the cylindrical surface enclosing the
egg, obtained by translating the curve representing the boundary of the egg
in the $h=0$ plane vertically upwards. Using the formula (\ref{diffintfl}),
we see that the first integral on the LHS, representing the area enclosed by
the projection of the trajectory onto the $\rho=0$ plane, remains positive
for all such trajectories, due to the clockwise flow of projections. As we
mentioned above, the second integral is always nonnegative, since the
integrand is. Hence we can show that all the arguments we derived for the
sourceless case extend to this case, again preventing favorable branch
changes from occurring. As no new pathologies appear, we conclude that the
no-go theorem must hold for this case too.

\section{The Higher Genus Corrections}

Here we shall consider the possibility of branch changing induced
by the string-loop corrections
when a stringy cosmological constant is present.
This corresponds to the action given by equation (\ref{hgcc}).
Equivalently, we will allow a general $\beta(\phi)$ and
constant $\Lambda$ in the equations of motion (\ref{e1}-\ref{e3}).
We will also briefly reflect on the case when $\Lambda$ is not
constant, as the situation then becomes a combination of the
two previous cases.

Our analysis is analogous to the genus-one case. We will again find
(+) and ($-$) branches and egg regions
where branch changing may take place.
However, in this case we will
find that all fixed points are located on the egg
boundaries, with attractors for both (+) and ($-$) branches
located on the upper surface of the egg.
By the weak coupling expansion
(\ref{bform}) we know that $B(\phi)$ is positive in a
large region of $\phi$, starting from $\phi
\rightarrow -\infty$. If we restrict
ourselves to eggs in this region,
we will show that the (+) branches originating
outside of the egg region cannot reach these fixed points, nor
can they use the eggs to change branches.
Allowing the conformal factor, $B(\phi)$, to become negative
will produce completely different
behavior.
In this case, we will see that we can construct completely
nonsingular cosmologies in the string frame, ending in a deSitter phase.
However, translating to the Einstein
frame we find such phase trajectories consist of two singular
Einstein branches separated by the point where $B(\phi)=0$,
where the conformal transformation to the Einstein frame is
singular.

Solving the constraint equation for $\dot \phi$ we find
that we should define the analog of the egg function (\ref{egg}) by:
\begin{eqnarray}
e=\sqrt{3(4-4 \beta+3 \beta^2) h^2+8 (1-\beta) \Lambda}
\label{pdegg}
\end{eqnarray}
Also, the equation for $\dot \phi$ is given by:
\begin{eqnarray}
\dot \phi=(\pm e+3 h (2-\beta))/(4 (1-\beta))
\label{pddotphi}
\end{eqnarray}
In the region where $B(\phi)>0$, the upper sign refers to the (+) and the lower
to the ($-$) branch. This ans\"atz is in accord with the
definition of the two branches in the previous section, as can be verified
by setting $\beta=0$. In the region where $B(\phi)<0$ we will
reverse this sign convention, the upper sign will refer to ($-$)
branch and the lower to (+). This curious reversal of our
conventions is needed to keep $\dot \phi$ continuous across
a sign change in $B(\phi)$ where $\beta$ becomes singular.
We will find continuous evolution through this line
(where $B=0$). To further
justify this convention, consider
the equations of motion in terms of the non-singular
quantities $B(\phi)$, $B'(\phi)$ and $B''(\phi)$. We see
that the reduction to the form (\ref{pddotphi}) requires
extracting the quantity $B(\phi)^2$ from the radical,
creating a sensitivity to the sign of $B(\phi)$.
The equation for $\dot h$ is complicated and will
not be needed here.

Next we examine the locations and character of fixed points.
Inserting $\dot \phi=\ddot \phi=\dot h=0$ into the equations
of motion we obtain the two conditions:
\begin{eqnarray}
0 &=& 3 h^2-\Lambda \\
\noalign{\medskip}
0 &=& (6 h^2-\Lambda) (\beta-2)=0
\end{eqnarray}
We notice that we need $\Lambda>0$ to get any
fixed points at all.  Furthermore, in this case a solution of these
equations is given by $3 h^2=\Lambda$
and $\beta=2$.
Putting this solution into (\ref{pdegg}) we find $e=0$,
and therefore all the fixed points are on the egg,
in contrast to the genus-zero case with dilaton self-interactions.

The condition $\beta=2$ corresponds
to $B'(\phi)=0$, so we see that the fixed points are extrema of $B(\phi)$.
As we noted before, this is to be expected from the
corresponding problem in the Einstein frame, which is given as a
dilaton evolution in a potential of the general form
$V(\phi)=\Lambda/B(\phi)$. We will use this fact
to classify the fixed points, as it will be difficult
to study their properties directly in the string frame.
Thus we see that positive
minima of $B(\phi)$ (equivalent to $\beta'>0$)
are maxima of $V(\phi)$,
and we conclude these are hyperbolic saddle points.
Positive maxima of $B(\phi)$ (equivalent to $\beta'<0$)
are positive minima of
$V(\phi)$, and thus these will be attractors for $h>0$
and repellers for $h<0$. We can easily extend these
results to regions where $B(\phi)<0$ by noting
that the sign of $B(\phi)$ does not enter into
the equations of motion, so we can simply invert
$B(\phi)$.

At this point, we need to investigate the shape and location of the egg.
Solving $e=0$ for $h^2$ we find:
\begin{eqnarray}
h^2=8 (\beta-1) \Lambda/(3(4-4 \beta+3 \beta^2))
\end{eqnarray}
Since the quadratic in the denominator is positive
definite, we find an egg boundary where $1 \le \beta \le \infty$.
Thus the egg begins at $h=0$ at a value of $\beta=1$, increases
to a peak of $h^2=\Lambda/3$ at $\beta=2$, and decreases
again to $h=0$ as $\beta \rightarrow \infty$.
Comparing
this with our results about the locations of the fixed
points, we conclude that the fixed points are
precisely at the peaks above and below the egg
(the topological saddle points of
the egg boundary).

Finally, we examine
the nature of curves along which $\dot \phi$ changes sign.
{}From (\ref{pddotphi}) we see that in the case where
$B(\phi)>0$, these are simply
the horizontal segments defined by $3 h^2-\Lambda=0$ with
$h (2-\beta)<0$ for the (+) branch and $h (2-\beta)>0$
for the ($-$) branch (these begin or end on one of
the aforementioned fixed points) and the vertical segments (where
$\dot \phi$ is singular) defined by $\beta=1$ with
$h>0$ for the (+) branch and $h<0$ for the ($-$) branch
(which begin at the ends of the egg). In the region
where $B(\phi)<0$, the same results hold with a
reversal of the branch designations. At the boundaries
between different signs of $B(\phi)$, we find
singular lines of $\dot \phi$ sign change. At
a line where $B(\phi)=0$, $B'(\phi)<0$ we find this discontinuity
for $h<0$ for (+) branch and $h>0$ for ($-$) branch,
and where $B'(\phi)>0$ we just reverse the branch labels.

Specializing to the case $B(\phi)>0$, marking each of these regions
with the sign of $\dot \phi$ flow gives
us the complete picture of $\phi$ flow in the phase
plane around a $B(\phi)>0$ egg, see Figs.~(5) and (6).
Consideration of the direction of flow alone leads us
to the no-go result in this case. We can easily see that
(+) branch solutions approaching the egg region from
the right or the left cannot get into the region
vertically above the egg where they could find an
attractor. Thus if they are to hit the egg they
must do it below and convert to ($-$).
But now we notice that a ($-$)
branch solution below the egg cannot emerge from
this region without converting to a (+). Combining
these two facts together, we see that the egg is
again unable to change a (+) branch to a ($-$) branch.
The case of a trajectory passing through an endpoint
of the egg is easily dealt with. Putting the conditions
characterizing the egg endpoints ($\beta=1$ and $h=0$)
into (\ref{e3}), we conclude $\Lambda=0$. Since we
are interested in the case $\Lambda>0$, the trajectories
do not flow through the endpoints.

If we begin with a $B(\phi)$ with
a negative minimum, we reach quite different conclusions
since the barriers surrounding attractors for (+) branch flow
have fallen.
Although one might suspect that all solutions beginning
in the region where $B(\phi)>0$ will not be able to cross
over a point where $B(\phi)=0$, this is not true. In the
string frame there exist solutions which cross this region
without hesitation, as numerical integration shows (see
Figs.~(7) and ~(8)).  In Fig.~(7), we show a solution which
is always undergoing expansion, while in Fig.~(8), we
show a solution which starts out in a contracting phase, undergoes
multiple branch changes and finishes in an expanding, asymptotically
deSitter state (as does the former solution). Both
of these solutions are non-singular in past as well
as in the future. These
solutions are quite peculiar in nature. It is actually quite
easy to connect our linear dilaton
solutions in the asymptotic past to deSitter
solutions in the asymptotic future, using these solutions.
The resulting configurations have several attractive features:
i) the evolution very naturally flows
to an attractor, ii) examining (\ref{bform}), we see that
it is much easier and more natural for the loop corrections
to have a negative minimum than a positive maximum
without introducing large unperturbative coefficients.
However, analyzed in the Einstein frame, the continuous string
frame evolution splits into two Einstein evolutions, the first
contracting to a singularity and the second expanding
out of a singularity. This is symptomatic of the fact that
the conformal transformation to the Einstein frame is ill-defined.
Even in the string frame, the apparent change of signature
is quite curious.  Though our future history is described by
a metric with a ($-$,+,+,+) signature, it begins with a
(+,$-$,$-$,$-$) signature, and because all of the terms in the Lagrangian
also change sign, the equations of motion are unaffected.
The curiosity occurs just at the point when B=0, where there is
no metric, and we are dealing with a topological field theory.
We hope to address this issue in more detail elsewhere.

In retrospect, we can now see that even a non-constant $\Lambda$ is extremely
unlikely to solve the graceful exit problem. Namely, the situation would
correspond to a combination of the two cases we developed the no-go theorems
for, i.e. the dilaton self interaction superimposed with the higher-genus
corrections. We would end up with eggs of both types considered, which
separately cannot facilitate a favorable branch change. A novelty would be
that the integral formula (\ref{diffint}) could not be given a simple
geometric interpretation for all cases. However,  for (+) branch solutions
extendible to linear dilaton vacua in the past, these aberrations would
typically be small, and should not induce any qualitatively new behavior.
Thus it appears that the only way to salvage the Pre-Big-Bang scenario
is to resort to the higher order terms in the $\alpha'$ expansion, with the
difficulties which this approach brings, as explained in section 4.

We close this section with the note that there still exists a possibility to
incorporate inflation in string theory using the model including
the higher genus corrections as described by
the action (\ref{hgcc}). Namely, it is not difficult to see that we can trap
a ($-$) branch originating away from the egg in a fixed point with $h>0$,
resulting in asymptotically deSitter inflation. This scenario was recently
analyzed in detail by Damour and Vilenkin \cite{dv},
in Einstein conformal frame.
We should only mention here that this scenario is very similar to
standard inflationary models, in the sense that the resulting universe starts
from a Big-Bang cosmological singularity.

\section{Conclusion}

We have derived an exact no-go theorem for string cosmology in
very general circumstances, ruling out the possibility of resolving the
graceful exit problem by branch changing catalyzed by dilaton
selfinteraction, fluid sources and higher genus corrections. Our analysis
was based on the investigation of phase space properties of the model,
resulting in precise and strong answers concerning the evolution of the
universe governed by the graviton-dilaton sector. We can still incorporate
inflation in the model using a combination of a nonzero stringy cosmological
constant with the higher-genus corrections, with the resulting cosmology
looking like a universe starting from a Big Bang singularity and
asymptotically evolving towards a deSitter phase, with a decoupled dilaton.

In addition, we have found a class of nonsingular solutions in the
string frame, for the case when the conformal coupling $B(\phi)$
becomes negative for some values of the dilaton.
These solutions evolve out of the linear dilaton vacua in the past
and asymptotically approach deSitter expansion in the future,  passing through
the value $B=0$ without hesitation. During this
evolution, however, the relative sign of the action changes, which
may require a topological description of the
Universe at the point where B=0.
We hope to address these solutions in more detail in the future.

Finally we observe that the only option still open for incorporating the
original Pre-Big-Bang scenario  is to resort to higher derivative
terms in the $\alpha'$ expansion. In this approach we must consider
systematically all the terms in the $\alpha'$ expansion, and this can only be
implemented via the exact conformal field theory construction. At this moment,
it appears that this goal is still beyond our means.

\vskip 2truecm

\noindent {\bf Acknowledgements}
\vskip 0.7truecm
We would like to thank B.A. Campbell, C. Kounnas and R.C. Myers for helpful
conversations. This work was supported in part by  DOE grant
DE-FG02-94ER40823, and in part by NSERC of Canada. NK was also supported
in part by an NSERC postdoctoral fellowship.

\newpage
\noindent{\bf{Figure Captions}}

\vskip.3truein

\begin{itemize}

\item[]
\begin{enumerate}
\item[]
\begin{enumerate}
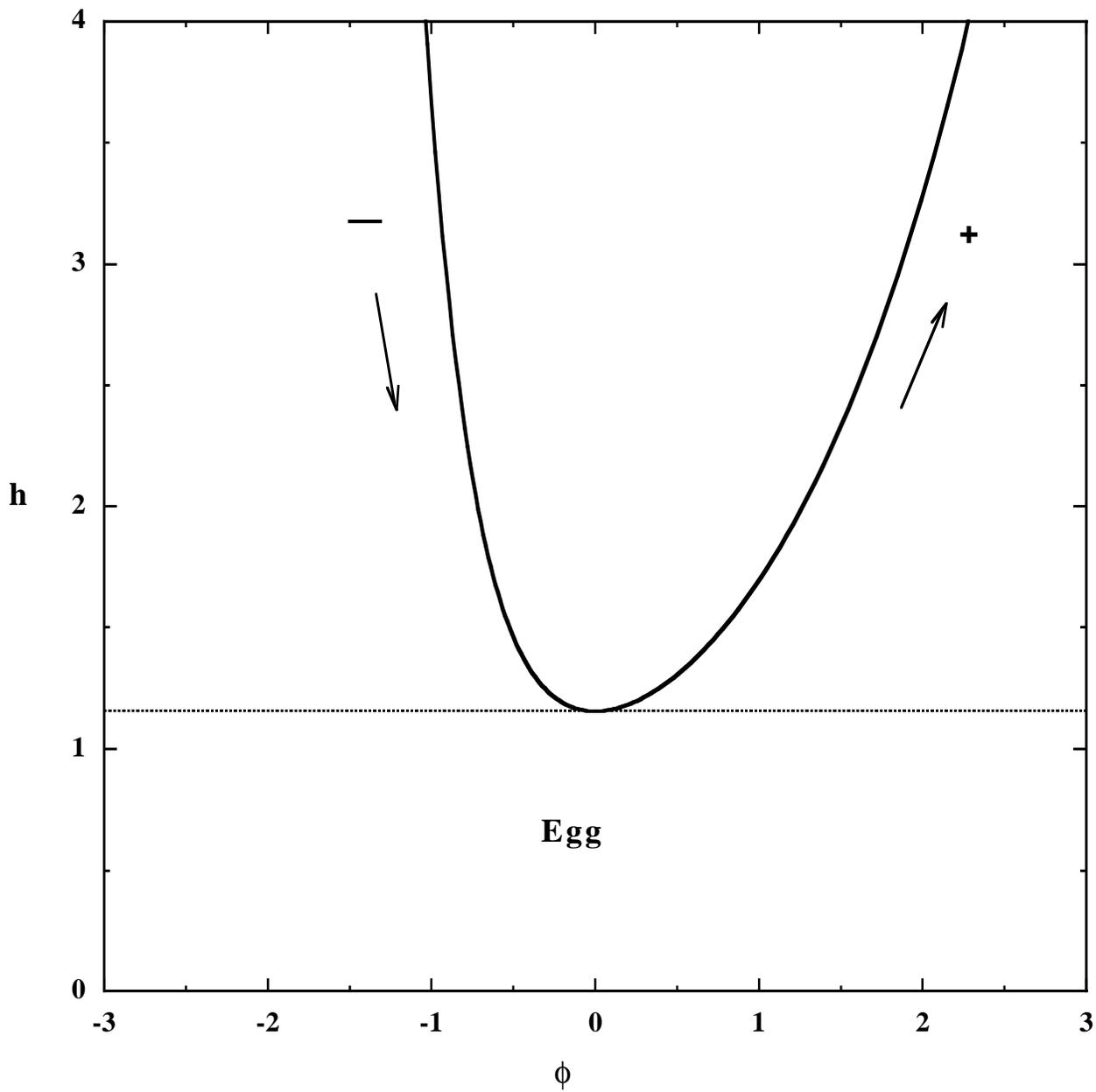
\item[{\bf Figure 1:}] Evolution in the case with a negative
cosmological constant ($\Lambda=-2$), with egg boundaries along the
line $h=\pm \sqrt{4/3}$. Since the branch change is in the wrong
direction ($(-) \rightarrow (+)$), this solutions has singularities
in both the past and the future.

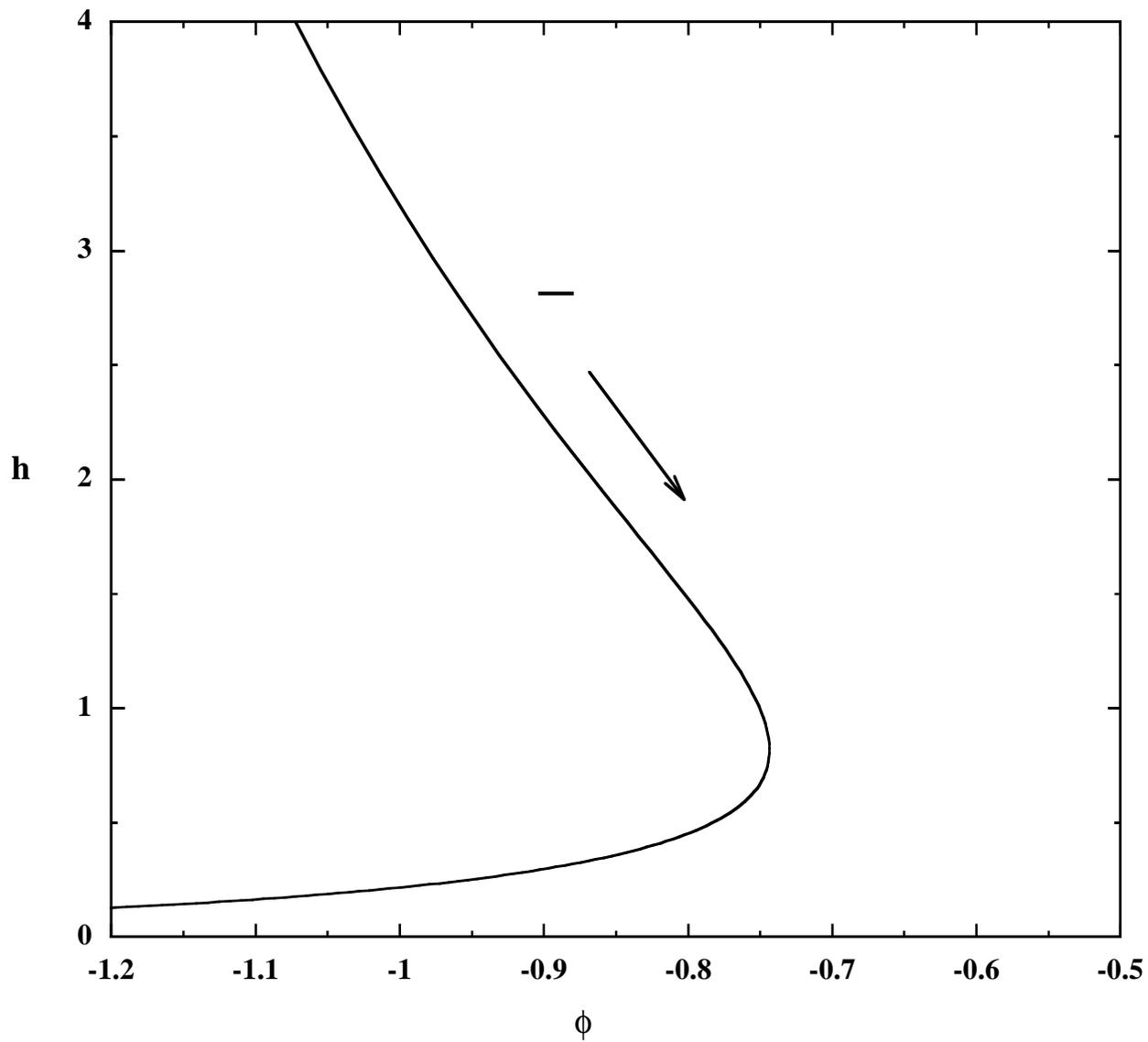
\item[{\bf Figure 2:}] A ($-$) branch solution with the positive
cosmological constant ($\Lambda=2$). The universe evolves from a past
singularity into a linear dilaton vacuum in the future.

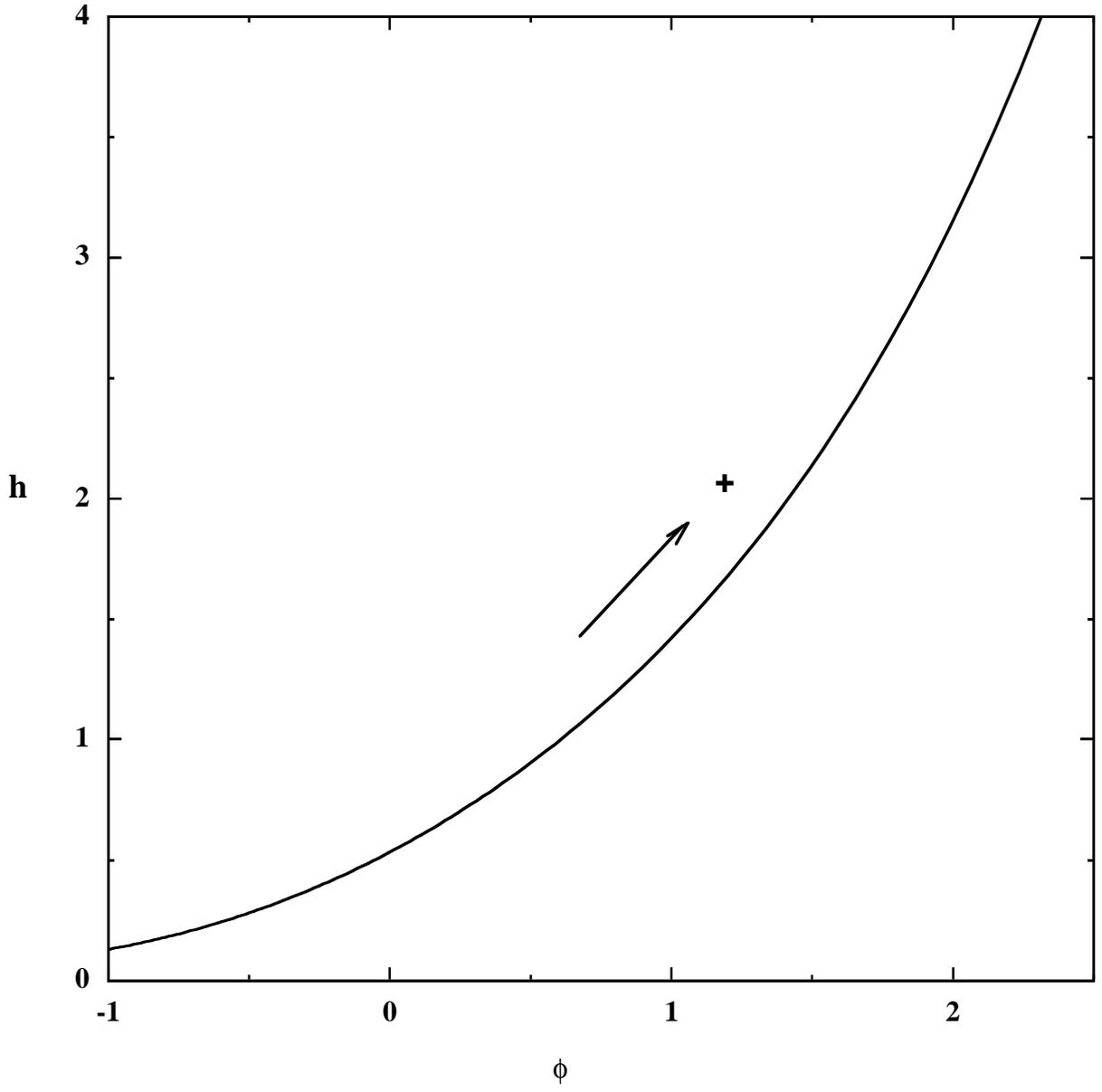
\item[{\bf  Figure 3:}] A ($+$) branch solution with the positive
cosmological constant ($\Lambda=2$). The universe evolves from a
linear dilaton vacuum into a future singularity.

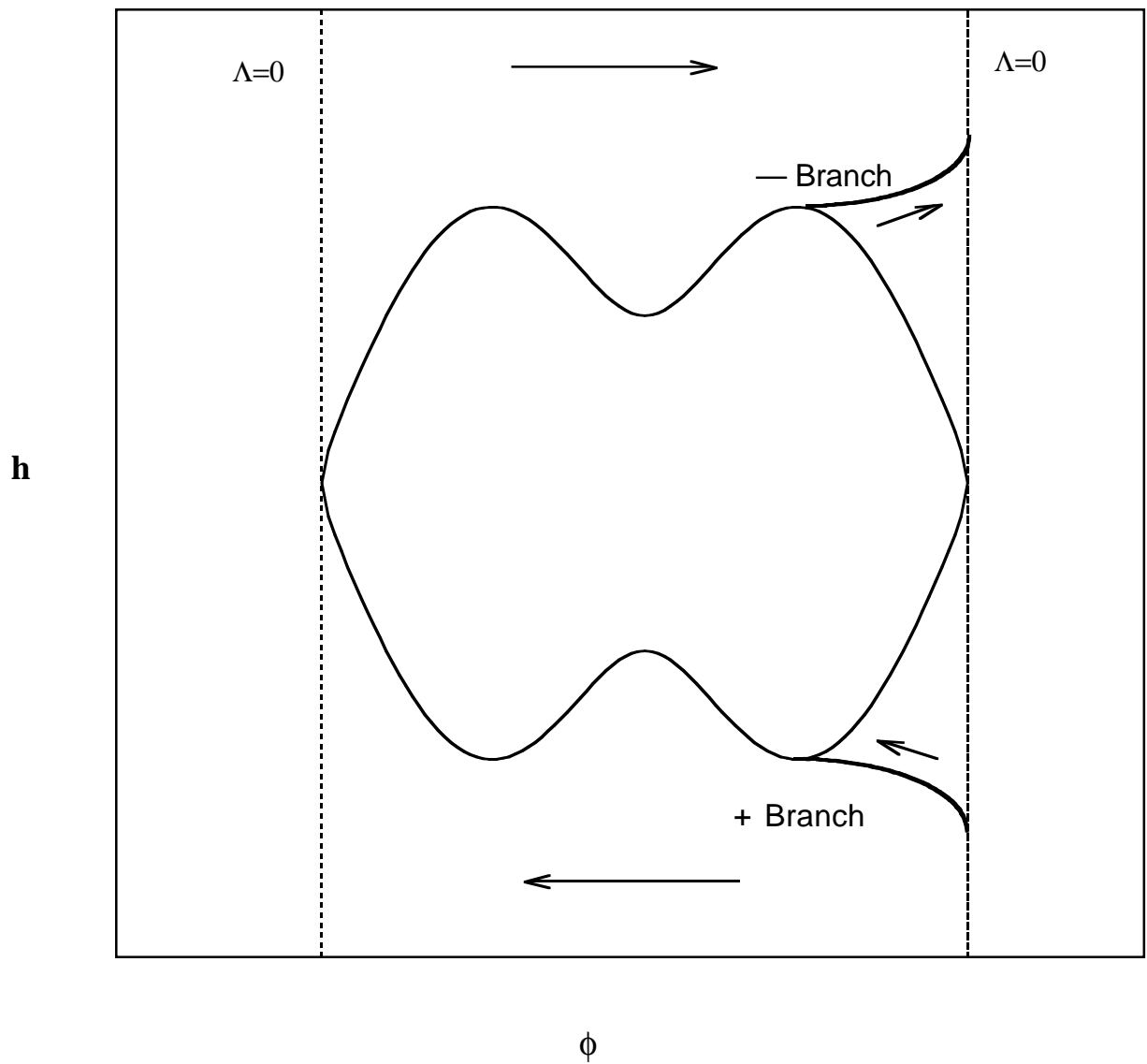
\item[{\bf  Figure 4:}]  Examples of trajectory segments ruled
out by the second and third arguments of the no-go theorem.
The ($-$) branch cannot leave the egg and exit to the right,
nor can the (+) branch enter from the right and hit the egg.
Here the vertical lines only demark the ends of the egg, and
are not the boundaries of different directions of $\phi$
flow.

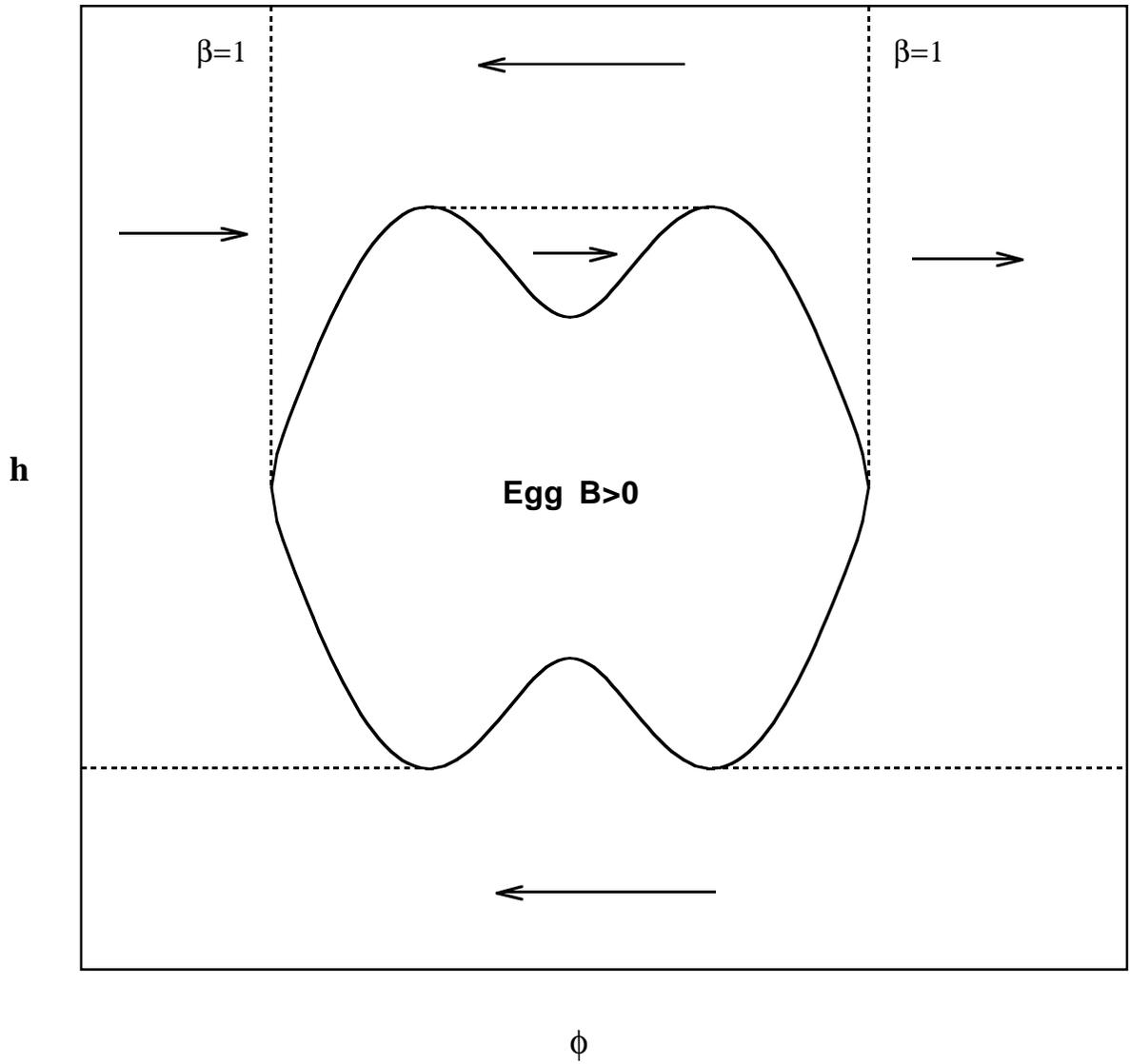
\item[{\bf  Figure 5:}] The boundaries of the regions of
uniform direction of $\phi$ flow of the (+) trajectories.
Notice that a (+) trajectory outside of the region above
the egg cannot enter the region where the deSitter
attractors are located. To get the corresponding picture
for the ($-$) branch trajectories, we need to invert this picture
and reverse the arrows.

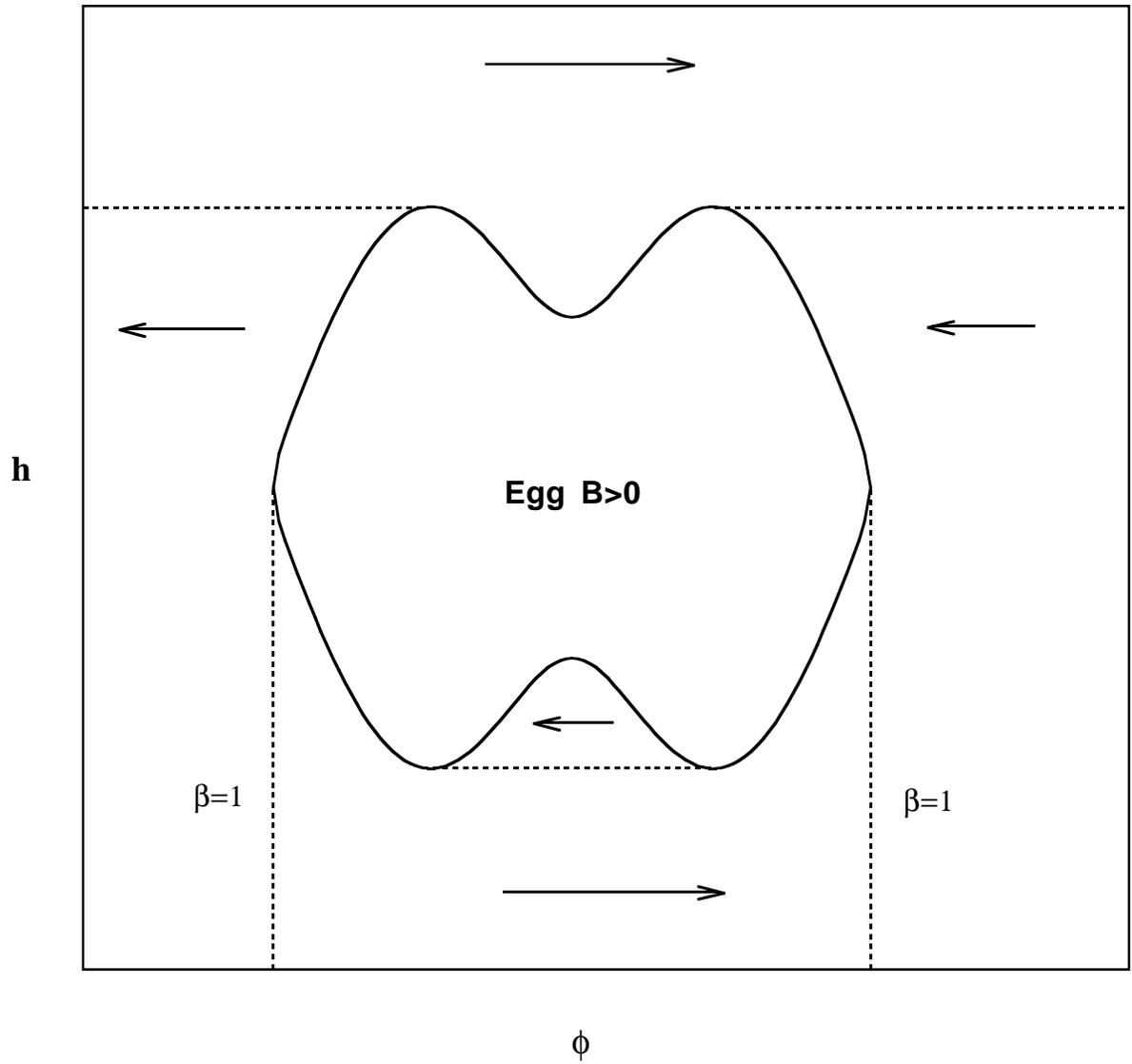
\item[{\bf  Figure 6:}] As in Figure 5 for the ($-$) branch.

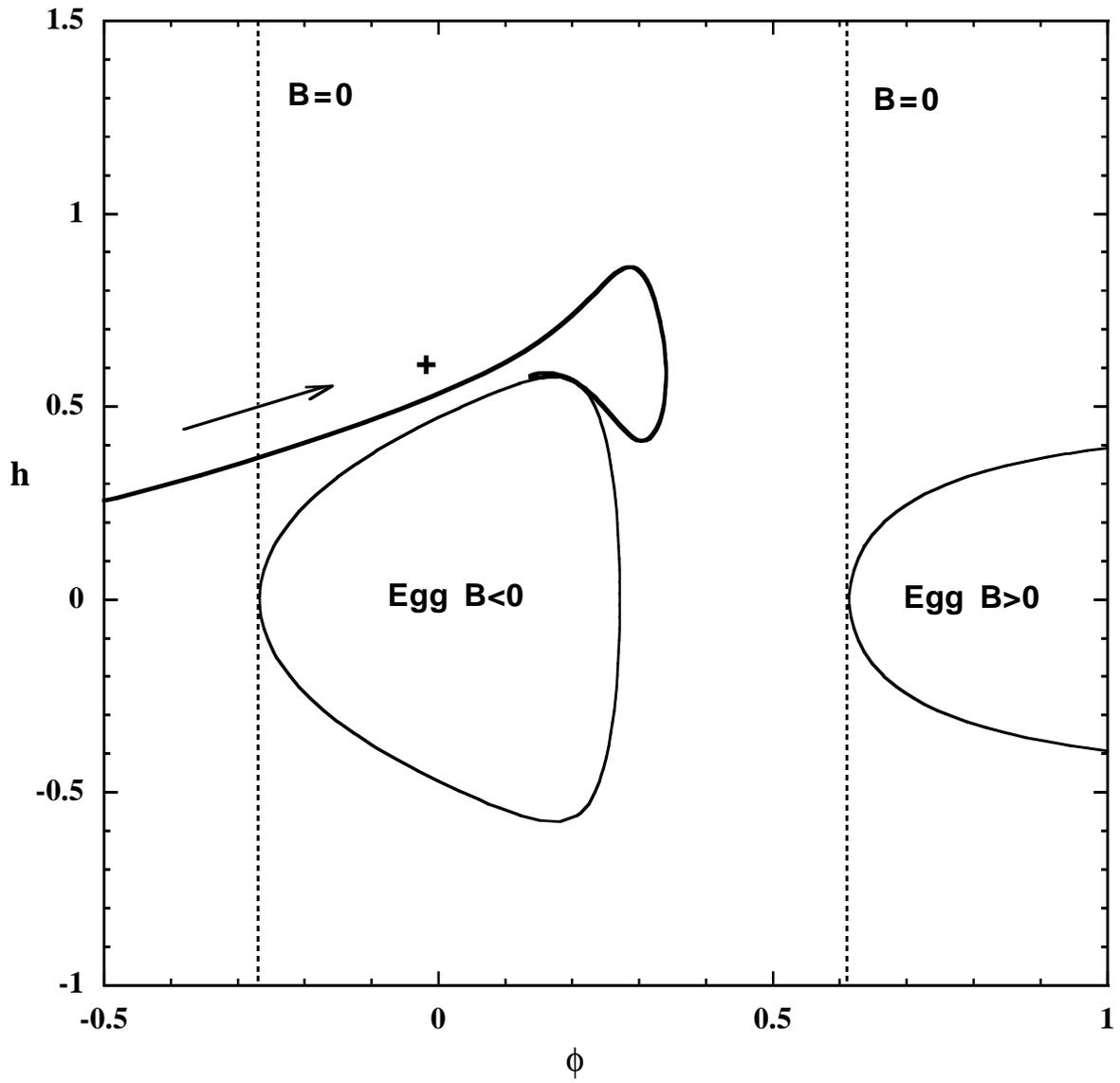
\item[{\bf  Figure 7:}] A nonsingular (+) branch trajectory
ending in a deSitter phase attractor on a $B<0$ egg,
$B(\phi)=e^{-2 \phi}-2+0.5e^{2 \phi}$ and $\Lambda=1$.

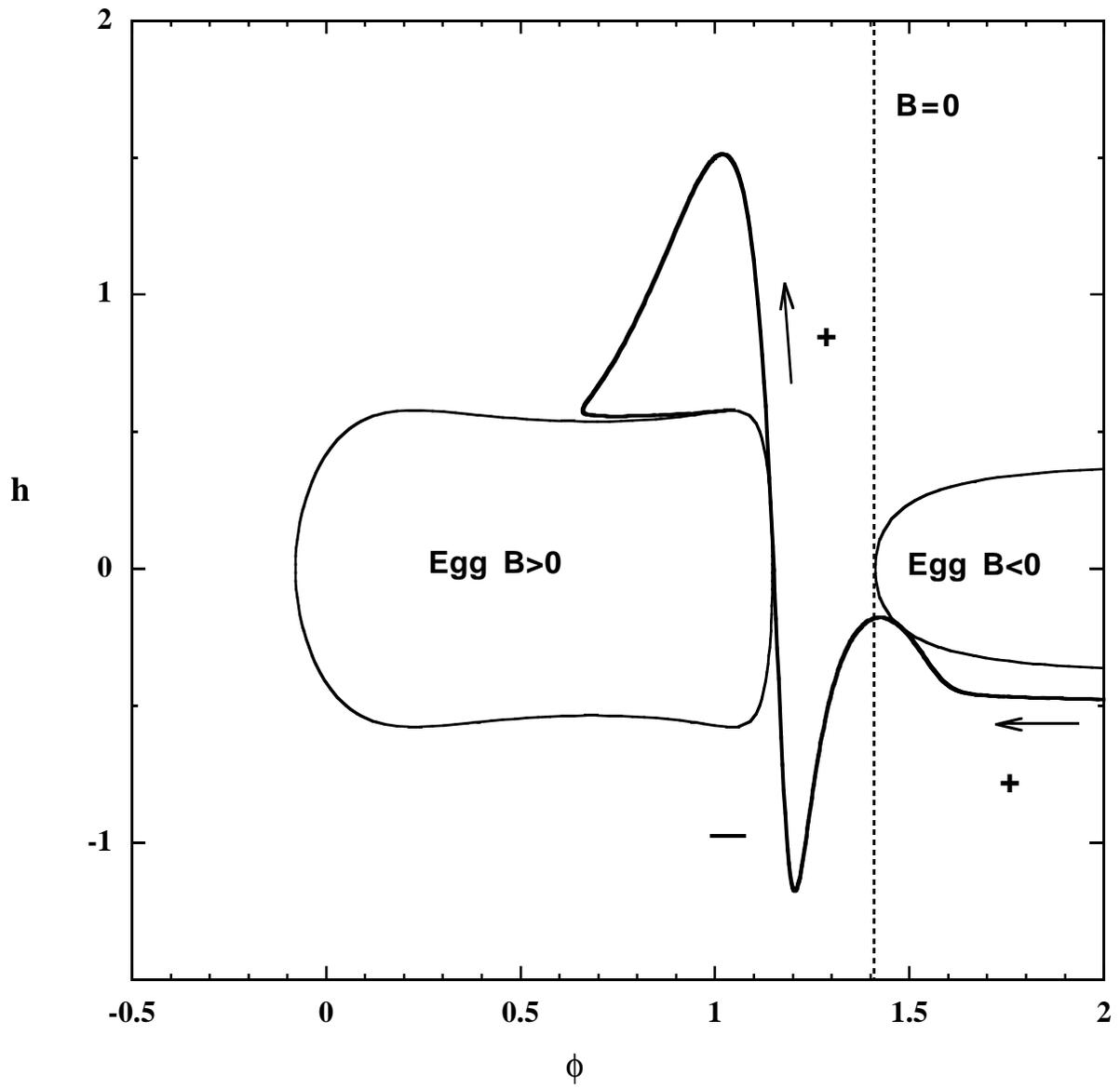
\item[{\bf  Figure 8:}] A nonsingular (+) branch trajectory
ends in a deSitter phase attractor on a $B>0$ egg. This
solution evades the terms of our no-go theorem by using
a $B<0$ egg in the strong coupling region to perform
the ($(+) \rightarrow (-)$) before the capture on the $B>0$
egg. $B(\phi)=e^{-2 \phi}+0.5e^{2 \phi}-0.03e^{4 \phi}$ and
$\Lambda=1$.

\end{enumerate}
\end{enumerate}
\end{itemize}
\end{document}